\documentclass[submission, Proceedings]{SciPost}

\hypersetup{
    colorlinks,
    linkcolor={red!50!black},
    citecolor={blue!50!black},
    urlcolor={blue!80!black}
}

\usepackage[bitstream-charter]{mathdesign}
\usepackage{subcaption}
\usepackage{epsfig}
\usepackage{pst-grad} 
\usepackage{pst-plot} 
\usepackage{physics}
\usepackage{tensor}
\usepackage{slashed}
\usepackage{graphicx}
\usepackage{simplewick}
\usepackage[normalem]{ulem}

\DeclareSymbolFont{usualmathcal}{OMS}{cmsy}{m}{n}
\DeclareSymbolFontAlphabet{\mathcal}{usualmathcal}

\newcommand{\un}[1]{\underline{#1}}

\newcommand{\tord}{\textrm{T} \:}
\newcommand{\atord}{\bar{\textrm{T}} \:}

\newcommand{\as}{\alpha_s}

\def\eq#1{{Eq.~(\ref{#1})}}
\def\fig#1{{Fig.~\ref{#1}}}

\begin{document}

\begin{center}{\Large \textbf{
Quark Sivers Function at Small-$x$: Leading contribution from the Spin-Dependent Odderon\\
}}\end{center}

\begin{center}
M. G. Santiago\textsuperscript{$\star$}
\end{center}

\begin{center}
 Department of Physics, The Ohio State University, Columbus, OH 43210, USA
\\

santiago.98@osu.edu
\end{center}

\begin{center}
\today
\end{center}

\definecolor{palegray}{gray}{0.95}
\begin{center}
\colorbox{palegray}{
  \begin{tabular}{rr}
  \begin{minipage}{0.1\textwidth}
    \includegraphics[width=22mm]{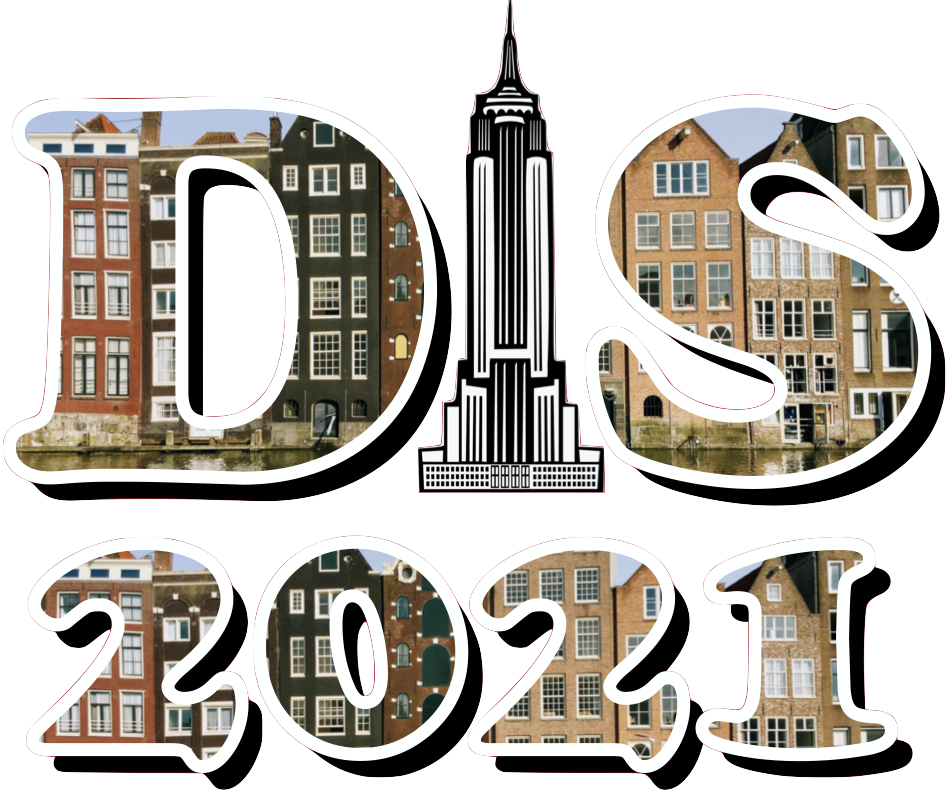}
  \end{minipage}
  &
  \begin{minipage}{0.75\textwidth}
    \begin{center}
    {\it Proceedings for the XXVIII International Workshop\\ on Deep-Inelastic Scattering and
Related Subjects,}\\
    {\it Stony Brook University, New York, USA, 12-16 April 2021} \\
    \doi{10.21468/SciPostPhysProc.?}\\
    \end{center}
  \end{minipage}
\end{tabular}
}
\end{center}

%%%%%%%%%%%%%%%%%%%%%%%%%%%%%%%%%%%%%%%%%%%%%%%%%%%%%%%%%%%%%%%%%%%%%%%%%%%%%%%%%%%%%%%%%%%%%

\section*{Abstract}
{\bf
We present the calculation of the leading contribution to the quark Sivers function at small-Bjorken $x$ as in \cite{kov}. This calculation uses the high energy scattering approximation and operator formalism developed in \cite{Kovchegov:2018znm,Kovchegov:2018zeq} to obtain a dominant contribution to the quark Sivers function coming from the spin-dependent odderon, in agreement with the results of \cite{Dong:2018wsp}. We then calculate this dominant contribution in the diquark model of the proton to obtain a small-$x$ estimate for the Sivers function.
}

%%%%%%%%%%%%%%%%%%%%%%%%%%%%%%%%%%%%%%%%%%%%%%%%%%%%%%%%%%%%%%%%%%%%%%%%%%%%%%%%%%%%%%%%%%%%%

\section{Introduction}
\label{sec:intro}
The Sivers function is an important object for understanding the transverse momentum dynamics of the proton in terms of it's constituent quarks and gluons. In particular, the quark Sivers function encodes information about the quark orbital angular momentum and the spin-orbit coupling which produces an asymmetric momentum distribution in the transverse plane. The small-Bjorken $x$ asymptotics of the analogous gluon Sivers function was found to be dominated by the spin-dependent odderon \cite{Boer:2015pni} in the eikonal approximation, and more recently it was shown that the quark Sivers function has the same dominant odderon contribution \cite{Dong:2018wsp}.  In light of the recent announcement by the D0 and TOTEM collaborations of the detection of the odderon in $pp$ and $p\overline{p}$ collisions \cite{Abazov:2020rus}, one might hope that the Sivers function will offer another venue to detect the odderon in future collider experiments. Here we present the calculation of the eikonal contribution to the quark Sivers function using the formalism developed in \cite{Kovchegov:2018znm,Kovchegov:2018zeq}, as well as an estimation of the spin-dependent odderon contribution to the Sivers function in the diquark model of the proton. This calculation is presented in detail and extended to sub-eikonal corrections in \cite{kov}.

%%%%%%%%%%%%%%%%%%%%%%%%%%%%%%%%%%%%%%%%%%%%%%%%%%%%%%%%%%%%%%%%%%%%%%%%%%%%%%%%%%%%%%%%%%%%%

\section{Quark Sivers Function in the Eikonal Approximation}
\label{sec:siveik}

\begin{figure}[ht]
\centering
\includegraphics[width=0.9\linewidth]{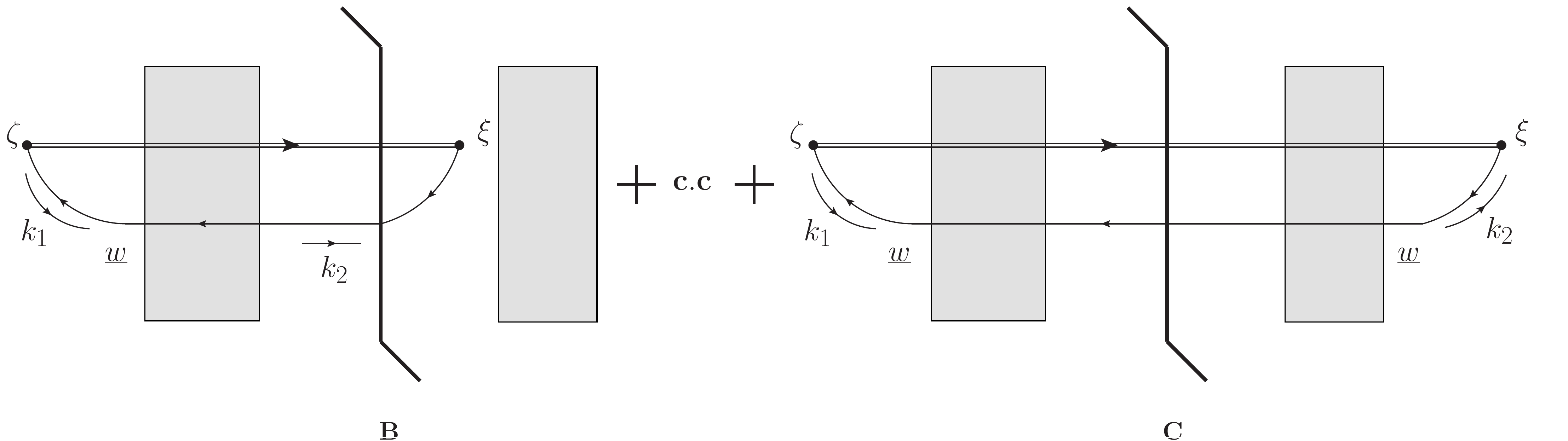}  
\caption{Diagrams of classes B and C with kinematics specified. In the diagram B the antiquark propagates from $\zeta$ with momentum $k_1$, undergoes a transverse spin-dependent interaction with the proton target at the transverse position $\un{w}$, then propagates to $\xi$ with momentum $k_2$. In the diagram C the anti-quark interacts with the shock wave again to the right of the cut, before arriving at the point $\xi$ with momentum $k_2$.}
\label{FIG:diagBCdet}
\end{figure}

Here we construct the small-$x$ asymptotics of the quark Sivers function. We begin with the operator definition for the unpolarized quark TMDs as functions of the longitudinal momentum fraction $x$ and transverse momentum $\underline{k}$ (cf. \cite{Meissner:2007rx})
\begin{equation}\label{S1}
f_1^q (x,k_T^2) - \frac{\un{k} \times \underline{S}_P}{M_P} f_{1 \: T}^{\perp \: q} (x,k_T^2) = \int \frac{\dd{r^-}\dd[2]{r_{\perp}}}{2 \, (2 \pi)^3} e^{i k \vdot r} \langle P, S | \bar{\psi}(0) \mathcal{U}[0,r] \frac{\gamma^+}{2} \psi(r) | P,S \rangle ,
\end{equation}
where $f_1^q$ is the unintegrated quark distribution, $f_{1 \, T}^{\perp \,q}$ the quark Sivers function, $M_P$ the proton mass, $\un{S}_P$ the proton spin, $k_T = |\un{k}|$, and $\mathcal{U}[0,r]$ the `staple' gauge link. Here we use the future pointing gauge link corresponding to SIDIS and take the proton to be moving in the light cone `$+$' direction with momentum $p_1^+$. We work in $A^- = 0$ light cone gauge of the projectile where the gauge link is a product of two straight, fundamental light-cone Wilson lines, $\mathcal{U}[0,r] = V_{\underline{0}}[0,\infty] V_{\underline{r}} [\infty,r^-]$, with light cone coordinates $x^{\pm} = t \pm z$,  $\un{x} = (x,y)$. These Wilson lines have the form
\begin{equation}
 V_{\un{x}} [\infty,a^-]=\mathcal{P}\textrm{exp}\Bigg[\frac{ig}{2}\int_{a^-}^{\infty}\dd{x}^- A^+ (x^-,\un{x})\Bigg],
\end{equation}
with $A^{\mu}$ the gluon field of the proton and $g$ the strong coupling constant. This operator product can be rewritten in the saturation/CGC formalism as
\begin{align}\label{S2}
f_1^q (x,k_T^2) - \frac{\un{k} \cross \underline{S}_P}{M_P} f_{1 \: T}^{\perp \: q} (x,k_T^2) = \frac{2 p_1^+}{2 (2 \pi)^3} \sum_X \int \dd{\xi^-} \dd[2]{\xi_{\perp}} \dd{\zeta^-} \dd[2]{\zeta_{\perp}} e^{i k \vdot (\zeta - \xi)}  \Big[\frac{\gamma^+}{2} \Big]_{\alpha \beta}  \\
\times  \Big{\langle} \bar{\psi}_{\alpha} (\xi) V_{\underline{\xi}} [\xi^-,\infty] | X \rangle \langle X | V_{\underline{\zeta}} [\infty, \zeta^-] \psi_{\beta} (\zeta) \Big{\rangle} \notag ,
\end{align} 
where we have inserted a complete set of states $\ket{X}$ between the Wilson lines and the angled brackets denote a semi-classical averaging of the operator over the wave function of the proton \cite{McLerran:1993ni,Kovchegov:2018znm}. Note that the Wilson lines are semi-infinite, connecting the quark fields at $\zeta$ and $\xi$ to infinity along the `$-$' direction. Taking the complete set of states as a final state cut, we can calculate the operator product diagrammatically, working to leading order in inverse powers of the center of mass energy for the scattering of a quark propagating through the proton (the eikonal approximation) and positive powers of $\as$, and show that the two contributing classes of diagrams are those in \fig{FIG:diagBCdet}, named classes B and C after the convention in \cite{Kovchegov:2018znm}. In these diagrams the thick horizontal lines represent the ordinary Wilson lines, the thin horizontal lines represent the propagating anti-quark with momentum $k_{1,2}$ and transverse position $\un{w}$, and the vertical lines represent the final state cuts. The propagating particle is an anti-quark as this is the only contributing final state $\ket{X}$ at leading order.

%%%%%%%%%%%%%%%%%%%%%%%%%%%%%%%%%%%%%%%%%%%%%%%%%%%%%%%%%%%%%%%%%%%%%%%%%%%%%%%%%%%%
%%%%%%%%%%%%%%%%%%%%%%%%%%%%%%%%%%%%%%%%%%%%%%%%%%%%%%%%%%%%%%%%%%%%%%%%%%%%%%%%%%%%

We  can calculate the contributions from diagrams in the classes B and C from \fig{FIG:diagBCdet} and obtain the eikonal contribution to the Sivers function as \cite{kov}
\begin{align}\label{ff1}
& \left[ f_1^q (x,k_T^2) -  \frac{\un{k} \cross \underline{S}_P}{M_P} f_{1 \: T}^{\perp \: q} (x,k_T^2) \right]_\textrm{eikonal} = \frac{4 p_1^+}{(2 \pi)^3} \int \dd[2]{\zeta_{\perp}}  \dd[2]{w_{\perp}} \frac{\dd[2]{k_{1 \perp}}\dd{k_1^-}}{(2\pi)^3} e^{i (\underline{k}_1 + \underline{k}) \vdot (\un{w} - \un{\zeta})}  \\
& \times \theta (k_1^-) \left\{ \frac{\un{k} \cdot \un{k}_1}{(x p_1^+ k_1^- + \underline{k}_1^2 ) (x p_1^+ k_1^- + \underline{k}^2)} \,  \Big{\langle} \tord \tr \left[ V_{\underline{\zeta}} \, V_{{\un w}}^\dagger \right] + \atord \tr \left[ V_{\underline{\zeta}} \, V_{{\un w}}^\dagger \right] \Big{\rangle} + \frac{\un{k}_1^2}{(x p_1^+ k_1^- + \underline{k}_1^2 )^2}  \, \Big{\langle} \tord \tr \left[ V_{\underline{\zeta}} \, V_{{\un w}}^\dagger \right]  \Big{\rangle}  \right\}. \notag
\end{align}
Note that we have inserted time ordering operators into the angle bracket correlators, and that these correlators are all traces of ordinary Wilson lines and are thus the familiar color dipole S-matrix. The Sivers function must be time-reversal odd, so we need to extract the term on the RHS of \eq{ff1} which changes sign under $\un{k} \rightarrow -\un{k}$. Taking $k_1 \rightarrow -k_1$, $\un{\zeta} \leftrightarrow \un{w}$ on the right-hand side of the expression, we can see that the $\un{k} \rightarrow -\un{k}$ sign change requires a term on the RHS which is odd under $\un{\zeta} \leftrightarrow \un{w}$. The dipole correlators can be decomposed \cite{Hatta:2005as,Kovchegov:2003dm} into a real Pomeron exchange $ \mathcal{S}_{\un{\zeta} \un{w}}$ which is symmetric under $\un{\zeta} \leftrightarrow \un{w}$ and an imaginary odderon exchange $ i \, \mathcal{O}_{\un{\zeta} \un{w}}$ which in antisymmetric as
\begin{equation}
\frac{1}{N_c} \Big{\langle}  \tord \tr [ V_{\underline{\zeta}}  V_{\underline{w}}^{\dagger }] \Big{\rangle} = \mathcal{S}_{\un{\zeta} \un{w}} + i \, \mathcal{O}_{\un{\zeta} \un{w}} ,
\end{equation}
so we conclude that the Sivers function comes from the odderon piece as
\begin{align}\label{siv}
 - \frac{\un{k} \cross \underline{S}_P}{M_P} f_{1 \: T}^{\perp \: q} (x,k_T^2) \Big|_\textrm{eikonal} & = \frac{4 i \, N_c \, p_1^+}{(2 \pi)^3} \int \dd[2]{\zeta_{\perp}}  \dd[2]{w_{\perp}} \frac{\dd[2]{k_{1 \perp}}\dd{k_1^-}}{(2\pi)^3} e^{i (\underline{k}_1 + \underline{k}) \vdot (\un{w} - \un{\zeta})} \theta (k_1^-) \\
& \times \left[ \frac{2 \, \un{k} \cdot \un{k}_1}{(x p_1^+ k_1^- + \underline{k}_1^2 ) (x p_1^+ k_1^- + \underline{k}^2)} + \frac{\un{k}_1^2}{(x p_1^+ k_1^- + \underline{k}_1^2 )^2}  \right] \, \mathcal{O}_{\un{\zeta} \un{w}}. \notag
\end{align}
This agrees with the spin-dependent odderon contribution as found in \cite{Dong:2018wsp}, which is proportional to $1/x$ and thus grows as we take smaller values of $x$. Small-$x$ evolution likely leaves this growth unaffected until the nonlinear saturation regime sets in, as the odderon is known to have an intercept equal to zero for linear evolution in various approximations for QCD \cite{Bartels:1999yt,Kovchegov:2012rz,Caron-Huot:2013fea}, and at strong coupling in $\mathcal{N}=4$ supersymmetric Yang-Mills theory \cite{Brower:2008cy}.

%%%%%%%%%%%%%%%%%%%%%%%%%%%%%%%%%%%%%%%%%%%%%%%%%%%%%%%%%%%%%%%%%%%%%%%%%%%%%%%%%%%%%%%%%%%%%

\section{Spin-Dependent Odderon Contribution in the Diquark Model}
\label{sec:diquark}

We need to establish that the odderon contribution to the Sivers function is nonzero. In \cite{Szymanowski:2016mbq} it was shown that the spin-dependent odderon can survive impact parameter integration and contribute to spin asymmetries using a diquark model calculation, assuming that the proton has an asymmetric parton distribution in the transverse plane. Here we will calculate the odderon contribution to the Sivers function using the same model, where the Lagrangian  has a spinor field $\psi_P$ for the (point-like) proton, the usual spinor fields for the quarks $\psi_q$, and a complex scalar diquark field $\varphi$ which has mass $M$ roughly equal to the proton mass $M \approx M_P$ and the color quantum numbers of an antiquark. The interaction term between these fields is a Yukawa coupling $\mathscr{L}_{int} = G \, \varphi^{* \, i} \, {\bar \psi}^i_q \, \psi_P$+c.c., with effective coupling constant $G$ for the splitting of the proton into a quark and a diquark. 

%%%%%%%%%%%%%%%%%%%%%%%%%%%%%%%%%%%%%%%%%%%%%%%%%%%%%%%%%%%%%%%%%%%%%%%%%%%%%%%%
\begin{figure}[h]
\begin{center}
\includegraphics[width= 0.5 \textwidth]{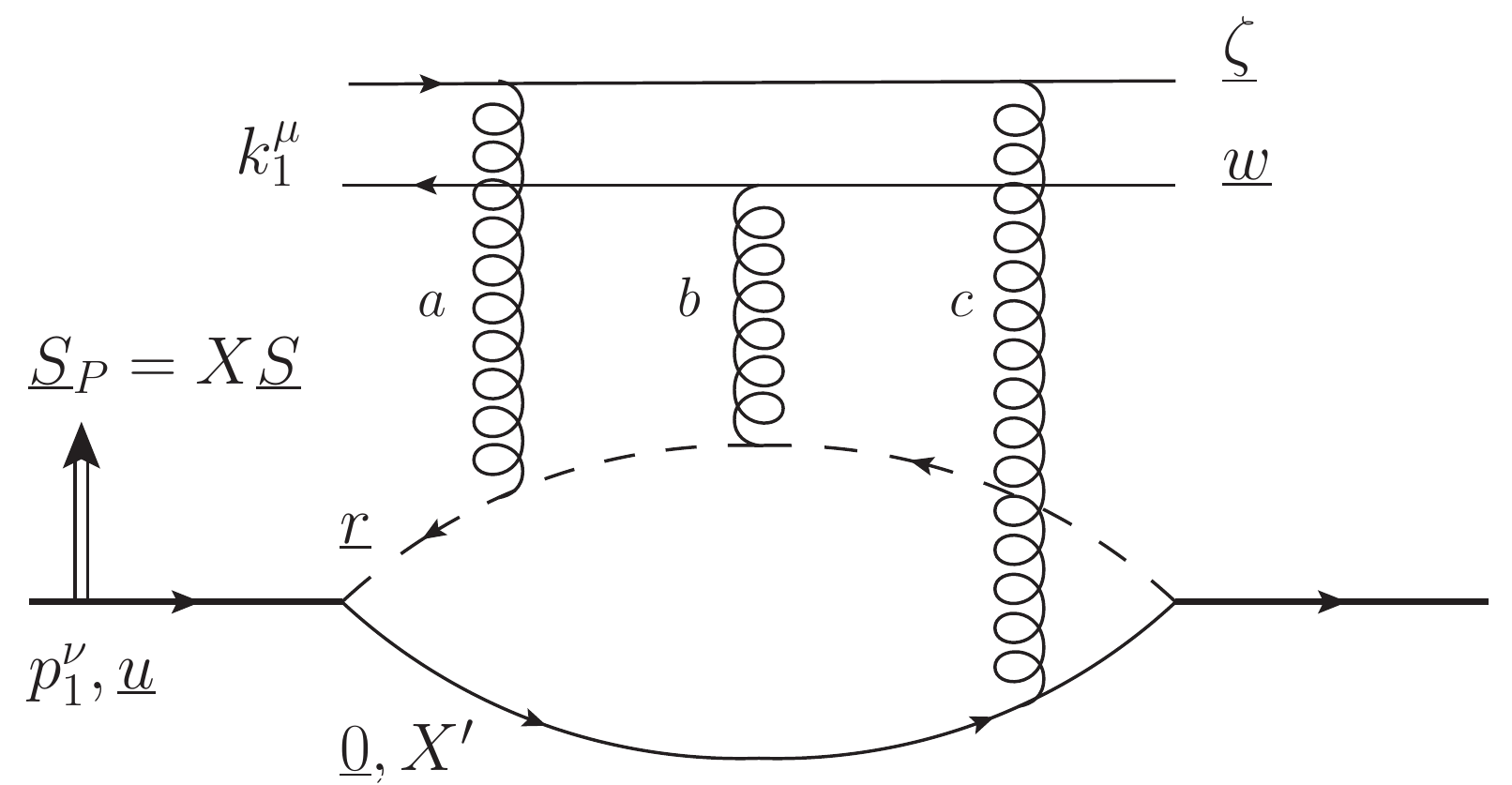} 
\caption{Diagram for the odderon exchange amplitude with the quark-diquark model of the proton, where all possible connections of the three gluons to the quark (solid line at the very bottom) and diquark (dashed line) and to the ${\un \zeta}, {\un w}$ dipole at the top should be summed over and the gluons are in the symmetric $d^{abc}$ color configuration.}
\label{FIG:odddiag}
\end{center}
\end{figure}
%%%%%%%%%%%%%%%%%%%%%%%%%%%%%%%%%%%%%%%%%%%%%%%%%%%%%%%%%%%%%%%%%%%%%%%%%%%%%%%%

We construct the odderon amplitude  $\mathcal{O}_{\un{\zeta} \un{w}}$ by calculating the diagram shown in \fig{FIG:odddiag}, where the proton splits into a quark-diquark color dipole and exchanges three gluons in the symmetric color configuration $d^{abc} = 2 \textrm{Tr} [ t^a \{t^b,t^c\}]$ with the quark-antiquark dipole at $\un{\zeta}$ and $\un{w}$. We need the light-cone wave function in the transverse spin basis for a proton at transverse position $\un u$ and with transverse polarization $X$, with the quark carrying the fraction $\gamma$ of the proton $p_1^+$ momentum and the transverse polarization $X'$ \cite{Kovchegov:2020kxg}
\begin{align}\label{coord_wf}
\psi_{X, X'} (\un{r}, \un{0}, \un{u}, \gamma)  = & \, \frac{G \tilde{m}_{\gamma} \sqrt{\gamma} (1 - \gamma )  }{2 \pi}  \delta^{(2)} \left( {\un r} - {\un u} - \gamma \, \un{r} \right) \\ 
& \times \left[\delta_{X, X'}  K_0 (\tilde{m}_{\gamma} r_{\perp})  - \frac{ i X  {r}^i }{ r_{\perp}}  K_1 (\tilde{m}_{\gamma} r_{\perp}) (i \delta_{X, X'} \delta^{i 2} - \delta_{X, - X'} \delta^{i 1} ) \right] , \notag
\end{align}
where $\tilde{m}_{\gamma}^2 =  (1-\gamma) m + \gamma M^2 - \gamma (1 - \gamma ) M_P^2 \approx \gamma^2 M_P^2$ in the limit of massless light quarks. We also need the amplitude for the triple gluon exchange \cite{Kovchegov:2012ga}
\begin{equation}\label{Ohat}
\hat{\mathcal{O}}_{\un{\zeta},\un{w}} = c_0 \, \alpha_s^3 \, \ln^3 \Bigg( \frac{|\un{\zeta} - \un{0} | \,  |\underline{w} - \underline{r}|}{|\un{w} - \un{0} | \, |\underline{\zeta} - \underline{r} | } \Bigg) ,
\end{equation}
with 
\begin{equation}\label{c0}
c_0 = - \frac{(N_c^2 - 4)(N_c^2-1)}{4N_c^3}.
\end{equation}
Colvoluting the triple gluon exchange amplitude \eq{Ohat} with the square of the light cone wave function \eq{coord_wf} yields the spin-dependent odderon amplitude, which we can insert in \eq{siv} to obtain 
\begin{align}\label{spinsiv}
 -& \frac{\un{k} \cross \underline{S}_P}{M_P} f_{1 \: T}^{\perp \: q} (x,k_T^2) \Big|_\textrm{eikonal} = \int\limits_0^1 \dd{\gamma} \frac{4 i N_c p_1^+ G^2  (1-\gamma) \tilde{m}_\gamma^2 c_0 \alpha_s^3}{(2 \pi)^6}  \int \dd[2]{r}_{\perp} \frac{\underline{S}_P \cross \underline{r}}{r_{\perp}}   \\
&\times K_0 (\tilde{m}_{\gamma} r_{\perp} ) K_1 ( \tilde{m}_{\gamma} r_{\perp} ) \int \dd[2]{\zeta}_{\perp} \dd[2]{w}_{\perp}  \int \frac{\dd[2]{k}_{1 \perp}\dd{k}_1^-}{(2\pi)^3} \theta (k_1^-)  e^{i (\underline{k}_1 + \underline{k} ) \vdot ( \underline{w} - \underline{\zeta})} \, \ln^3 \Bigg( \frac{\zeta_{\perp} | \underline{w} - \underline{r} |}{w_{\perp} | \underline{\zeta} - \underline{r} |} \Bigg) \notag \\
& \left[ \frac{2 \, \un{k} \cdot \un{k}_1}{(x p_1^+ k_1^- + \underline{k}_1^2 ) (x p_1^+ k_1^- + \underline{k}^2)} + \frac{\un{k}_1^2}{(x p_1^+ k_1^- + \underline{k}_1^2 )^2}  \right] , \notag
\end{align}
where the factor of $\underline{S}_P$ comes from the spin quantization axis $\underline{S}$ multiplied by the proton polarization $X$. We can evaluate \eq{spinsiv} using several approximations as in \cite{kov}, yielding the estimate 
\begin{equation}\label{oddsiv}
f_{1 \: T}^{\perp \: q}  (x,k_T^2)  \Big|_\textrm{eikonal}   =  \frac{1}{x}  \frac{N_c G^2 \, c_0 \alpha_s^3}{2 (2 \pi)^{5}} \, \frac{M_P^2}{\un{k}^2 \Lambda^2} ,
\end{equation}
where $\Lambda$ is in infrared cutoff. We have obtained a nonzero, eikonal contribution to the Sivers function from the spin-dependent odderon, similar to the results of \cite{Szymanowski:2016mbq}. We note that this result has interesting behavior  in the $\Lambda_{QCD},M_P \rightarrow 0$ limit. Since the IR cutoff $\Lambda$ must be proportional to $\Lambda_{QCD}$, we conclude that $\Lambda \sim \Lambda_{QCD} \sim M_P$. Therefore, $M_P^2/\Lambda^2$ ratio will remain constant in the $\Lambda_{QCD},M_P \rightarrow 0$ limit and the Sivers function \eq{oddsiv} will not vanish in the limit of zero proton mass. This may be a feature of using the quark--diquark model of the proton, but it calls for further analysis.

%%%%%%%%%%%%%%%%%%%%%%%%%%%%%%%%%%%%%%%%%%%%%%%%%%%%%%%%%%%%%%%%%%%%%%%%%%%%%%%%%%%%%%%%%%%%%

\section{Conclusion}
\label{sec:conc}

We have calculated the leading order contribution to the quark Sivers function at small-$x$, obtaining a contribution from the spin-dependent odderon as in \cite{Dong:2018wsp}. This is an eikonal contribution, which grows as $1/x$ as one probes the Sivers function at smaller and smaller values of $x$. This means that spin asymmetries generated by this eikonal contribution which are proportional to $xf_{1T}^{\perp q}$ will be constant, providing a possible route for detecting the odderon in single spin asymmetries such as in SIDIS at a future collider like the EIC. 

This calculation is shown in detail in \cite{kov}, where we expand upon this result and calculate the sub-eikonal contribution to the quark Sivers function and derive and solve its small-x evolution equation. This is accomplished by first calculating a general 'polarized Wilson line' operator which gives the couplings of the S-matrix for a quark with arbitrary polarization scattering off the background fields of a target proton to the arbitrary polarization of the target proton. The operator is comprised of sub-eikonal and sub-sub-eikonal corrections to the usual eikonal Wilson lines, and can in principle be used to construct the small-$x$ asymptotics of all eight leading-twist quark TMDs.

%%%%%%%%%%%%%%%%%%%%%%%%%%%%%%%%%%%%%%%%%%%%%%%%%%%%%%%%%%%%%%%%%%%%%%%%%%%%%%%%%%%%%%%%%%%%%

\section*{Acknowledgements}
\label{sec:acknowledgement}

We would like to thank Markus Diehl, Daniel Pitonyak and Jian Zhou for discussions. This material is based upon work supported by the U.S. Department of Energy, Office of Science, Office of Nuclear Physics under Award Number DE-SC0004286.

%%%%%%%%%%%%%%%%%%%%%%%%%%%%%%%%%%%%%%%%%%%%%%%%%%%%%%%%%%%%%%%%%%%%%%%%%%%%%%%%%%%%%%%%%%%%%

%%%%%%%%%%%%%%%%%%%%%%%%%%%%%%%%%%%%%%%%%%%%%%%%%%%%%%%%%%%%%%%%%%%%%%%%%%%%%%%%%%%%%%%%%%%%%

\nolinenumbers

\end{document}